\documentclass[preprint,showpacs,preprintnumbers,amsmath,amssymb]{revtex4}

\usepackage{epsfig}
\usepackage{graphicx}
\usepackage{dcolumn}
\usepackage{bm}

\newcommand{\ord}{{\cal O}}
\def\beq{\begin{equation}}
\def\eeq#1{\label{#1}\end{equation}}
\def\eeqn{\end{equation}}
\newcommand\iden{\leavevmode\hbox{\small1\normalsize\kern-.33em1}}

\let\jnfont=\rm
\def\RMP#1,{{\jnfont Rev.\ Mod.\ Phys }{\bf #1},}
\def\NPB#1,{{\jnfont Nucl.\ Phys.\ B }{\bf #1},}
\def\PLB#1,{{\jnfont Phys.\ Lett.\ B }{\bf #1},}
\def\EPJC#1,{{\jnfont Eur.\ Phys.\ Jour.\ C }{\bf #1},}
\def\PRD#1,{{\jnfont Phys.\ Rev.\ D }{\bf #1},}
\def\PRL#1,{{\jnfont Phys.\ Rev.\ Lett.\ }{\bf #1},}
\def\MPLA#1,{{\jnfont Mod.\ Phys.\ Lett.\ A }{\bf #1},}
\def\JPG#1,{{\jnfont J.\ Phys.\ G }{\bf #1},}
\def\CTP#1,{{\jnfont Commun.\ Theor.\ Phys.\ }{\bf #1},}
\def\JHEP#1,{{\jnfont JHEP \ }{\bf #1},}
\def\NPPS#1,{{\jnfont Nucl.\ Phys.\ Proc.\ Suppl.\ }{\bf #1},}
\def\CPC#1,{{\jnfont Computl.\ Phys.\ Commun.\ }{\bf #1},}

\def\o_slash{\not{\hbox{\kern-2.1pt $p_1$}}}
\def\p_slash{\not{\hbox{\kern-4.0pt $p_2$}}}
\def\q_slash{\not{\hbox{\kern-2.1pt $q_2$}}}
\def\e_slash{\not{\hbox{\kern-1.5pt $\epsilon_1^*$}}}
\def\f_slash{\not{\hbox{\kern-1.5pt $\epsilon_2^*$}}}

\newcommand{\be}{\begin{equation}}
\newcommand{\ee}{\end{equation}}
\newcommand{\bea}{\begin{eqnarray}}
\newcommand{\eea}{\end{eqnarray}}
\begin{document}

\preprint{\parbox{1.2in}{\noindent ~}}

\title{\ \\[10mm] Top quark FCNC decays and productions at LHC\\
                  in littlest Higgs model with T-parity}

\author{Xiao-Fang Han, Lei Wang, Jin Min Yang }

\affiliation{Key Laboratory of Frontiers in Theoretical
             Physics,Institute of Theoretical Physics, Academia Sinica,
             Beijing 100190, China \vspace*{1.5cm}}

\begin{abstract}
In the littlest Higgs model with T-parity (LHT) the newly
introduced mirror quarks have flavor-changing couplings with
the Standard Model (SM) quarks and may enhance the
flavor-changing neutral-current (FCNC) top quark interactions
which are extremely suppressed in the SM.
In this work we perform a comprehensive study for the
contributions of these mirror fermions to various top quark FCNC
decays and productions at the LHC, which
includes the decays $t \to c V (V=g, \gamma, Z)$,
$t \to cgg$ and the productions proceeding through the parton
processes $cg \to t$, $gg \to t \bar c$, $cg \to tg$, $cg \to
t\gamma$ and $cg \to t Z$. We find that although these
FCNC processes can be greatly enhanced by the LHT contributions,
they are hardly accessible at the LHC. Therefore, the LHT model
may not cause the FCNC problem in the top quark sector if the
top quark property is proved to be SM-like at the LHC.
\vspace*{1cm}
\end{abstract}


\maketitle

\section{Introduction}
As a possible solution to the hierarchy problem,
the little Higgs theory was proposed \cite{ref1}
and so far remains a popular candidate for new physics
beyond the Standard Model (SM). The littlest Higgs model
\cite{ref2} is a cute economical implementation of the little Higgs
idea, but is found to be subject to strong constraints from
electroweak precision tests \cite{ref3}, which would require raising
the mass scale of the new particles to far above TeV scale and thus
reintroduce the fine-tuning in the Higgs potential \cite{ref4}. To
tackle this problem, a discrete symmetry called T-parity is proposed
\cite{ref5}, which forbids the tree-level contributions from the
heavy gauge bosons to the observables involving only SM particles as
external states.
With the running of the LHC, these little Higgs models
will soon be put to the test.
To unravel the hints of these models,
the Higgs boson processes may be of primary importance because
these models significantly alter the
property of the Higgs boson \cite{LHT-higgs}.

Another sensitive probe for new physics like these little
Higgs models is the top quark processes.
As the heaviest known elementary particle, top quark may be
a window to look into the TeV-scale physics.
So far the top quark  properties have not been precisely
measured at the Tevatron collider due to the small statistics
and there remains plenty of room for new physics in the
top quark sector.  As a top quark factory, the LHC will allow to scrutinize
the top quark nature, which may provide clues to new physics \cite{topreview}.
For the little Higgs model with T-parity (LHT), one aspect of its phenomenology
in top quark sector is that the newly introduced mirror quarks
have flavor-changing couplings with the SM quarks
and may enhance the flavor-changing neutral-current (FCNC)
top quark interactions which are extremely suppressed in the SM \cite{tcvh-sm}.
Just like their effects in the rare decays of $K$ and $B$ mesons
\cite{blht1,blht2,blht3} as well as in the rare decays of the Higgs and
$Z$ bosons \cite{zsblht}, their contributions
to various top quark FCNC decays and productions at the LHC
may be significant and should be seriously checked.

In this work we collectively study the LHT contributions to the top
quark FCNC decays and productions at the LHC, which includes the
decay modes $t \to c V (V=g, \gamma, Z)$, $t \to cgg$ and the
productions proceeding through the parton processes $cg \to t$, $gg
\to t \bar c$, $cg \to tg$, $cg \to t\gamma$ and $cg \to t Z$. Some
of these processes have been studied in the literature
\cite{houhs,wangxl}, while the decay $t \to cgg$ and the production
$cg \to t$ have not yet been considered. As found in other new
physics models, like the supersymmetric models \cite{tcv-mssm} and
the technicolor models \cite{tcv-tc2}, these two channels have the
largest rates among the FCNC top quark processes. On the other hand,
the contributions of box diagrams, which are not included in the
calculations in the literature, should also be considered because
their contributions to the productions are at the same order as the
vertex loops. Further, since all these decays and productions depend
on a same set of parameters and are strongly correlated, they should
be studied and displayed collectively and comparatively.

 The work is organized as follows. In Sec. II we recapitulate the LHT model and
discuss the new flavor violating interactions which will contribute to
the FCNC processes considered in this work. In Sec. III we calculate
the LHT contributions to the top quark FCNC processes and present
some numerical results.
Finally, we give our conclusions in Sec. IV.

\section{The littlest Higgs model with T-parity}

The LHT model \cite{ref5} is based on a non-linear sigma model
describing the spontaneous breaking of a global $SU(5)$ down to a
global $SO(5)$ by a 5$\times$5  symmetric tensor at the scale
$f\sim\ord({\rm TeV})$. From the $SU(5)/SO(5)$ breaking, there arise 14
Goldstone bosons which are described by the "pion" matrix $\Pi$,
given explicitly by
\small
\be \label{Pi}
 \addtolength{\arraycolsep}{3pt}\renewcommand{\arraystretch}{1.3}
 \Pi=\left(\begin{array}{ccccc}
-\frac{\omega^0}{2}-\frac{\eta}{\sqrt{20}} &
-\frac{\omega^+}{\sqrt{2}} &
  -i\frac{\pi^+}{\sqrt{2}} & -i\phi^{++} & -i\frac{\phi^+}{\sqrt{2}}\\
-\frac{\omega^-}{\sqrt{2}} &
\frac{\omega^0}{2}-\frac{\eta}{\sqrt{20}} &
\frac{v+h+i\pi^0}{2} & -i\frac{\phi^+}{\sqrt{2}} & \frac{-i\phi^0+\phi^P}{\sqrt{2}}\\
i\frac{\pi^-}{\sqrt{2}} & \frac{v+h-i\pi^0}{2} &\sqrt{4/5}\eta &
-i\frac{\pi^+}{\sqrt{2}} & \frac{v+h+i\pi^0}{2}\\
i\phi^{--} & i\frac{\phi^-}{\sqrt{2}} & i\frac{\pi^-}{\sqrt{2}} &
-\frac{\omega^0}{2}-\frac{\eta}{\sqrt{20}} & -\frac{\omega^-}{\sqrt{2}}\\
i\frac{\phi^-}{\sqrt{2}} &  \frac{i\phi^0+\phi^P}{\sqrt{2}} &
\frac{v+h-i\pi^0}{2} & -\frac{\omega^+}{\sqrt{2}} &
\frac{\omega^0}{2}-\frac{\eta}{\sqrt{20}}
\end{array}\right).
\ee
\normalsize
Under T-parity the SM Higgs doublet $ H= \left(-i
\pi^+/ \sqrt{2}, (v+h+i\pi^0)/2 \right)^T$ is T-even while
other fields are T-odd. A subgroup $[SU(2)\times
U(1)]_{1}\times[SU(2)\times U(1)]_{2}$ of the $SU(5)$ is gauged and
at the scale $f$ it is broken into the SM electroweak symmetry
$SU(2)_L\times U(1)_Y$. The Goldstone bosons $\omega^{0}$,
$\omega^{\pm}$ and $\eta$ are respectively eaten by the new T-odd
gauge bosons $Z_{H}$, $W_{H}$ and $A_{H}$, which obtain masses at
$\ord(v^2/f^2)$ \be M_{W_H}=
M_{Z_H}=fg\left(1-\frac{v^2}{8f^2}\right), ~~ M_{A_H}=\frac{f
g'}{\sqrt{5}}\left(1-\frac{5v^2}{8f^2}\right), \ee with $g$ and
$g^\prime$ being the SM $SU(2)$ and $U(1)$ gauge couplings,
respectively.

The  Goldstone bosons $\pi^{0}$ and $\pi^{\pm}$ are eaten by the
T-even $Z$ and $W$ bosons of the SM, which obtain masses at $\ord(v^2/f^2)$
\be
M_{W_L}=\frac{gv}{2}\left(1-\frac{v^2}{12f^2}\right),\quad
M_{Z_L}=\frac{gv}{2\cos\theta_W}\left(1-\frac{v^2}{12f^2}\right).
\ee
The photon $A_L$ is also T-even and remains massless.

For each SM quark, a copy of mirror quark with T-odd quantum number
is added in order to preserve the T-parity. We denote them by
$u_H^i$ and $d_H^i$, where $i=1, 2, 3$ are the generation index.
In $\ord(v^2/f^2)$ their masses are given by
\be
m_{d_H^i}=\sqrt{2}\kappa_{q^i}f,\qquad
m_{u_H^i}=m_{d_H^i}(1-\frac{v^2}{8f^2}), \label{eq4}
\ee
where $\kappa_{q^i}$ are the diagonalized Yukawa couplings of the mirror
quarks.

Note that new flavor interactions arise between the mirror fermions
and the SM fermions, mediated by the T-odd gauge bosons or T-odd
Goldstone bosons. In general, besides the charged-current
flavor-changing interactions, the FCNC interactions between the
mirror fermions and the SM fermions can also arise from the mismatch
of rotation matrices. For example, there exist FCNC interactions
between the mirror up-type (down-type) quarks and the SM up-type
(down-type) quarks, where the mismatched mixing matrix is denoted by
$V_{H_{u}}$ ($V_{H_{d}}$) with $V^{\dag}_{H_{u}}V_{H_{d}}=V_{CKM}$.
We follow \cite{new-add} to parameterize $V_{H_{d}}$ with three
angles $\theta_{12}^d,\theta_{23}^d,\theta_{13}^d$ and three phases
$\delta_{12}^d,\delta_{23}^d,\delta_{13}^d$ \small \be
\left(\begin{array}{ccc} c_{12}^d c_{13}^d & s_{12}^d c_{13}^d
e^{-i\delta^d_{12}}& s_{13}^d e^{-i\delta^d_{13}}\\
-s_{12}^d c_{23}^d e^{i\delta^d_{12}}-c_{12}^d s_{23}^ds_{13}^d
e^{i(\delta^d_{13}-\delta^d_{23})} & c_{12}^d c_{23}^d-s_{12}^d
s_{23}^ds_{13}^d e^{i(\delta^d_{13}-\delta^d_{12}-\delta^d_{23})} &
s_{23}^dc_{13}^d e^{-i\delta^d_{23}}\\
s_{12}^d s_{23}^d e^{i(\delta^d_{12}+\delta^d_{23})}-c_{12}^d
c_{23}^ds_{13}^d e^{i\delta^d_{13}} & -c_{12}^d s_{23}^d
e^{i\delta^d_{23}}-s_{12}^d c_{23}^d s_{13}^d
e^{i(\delta^d_{13}-\delta^d_{12})} & c_{23}^d c_{13}^d
\end{array}\right).
\ee
\normalsize

\section{FCNC top quark processes in the LHT model}
\begin{figure}[htb]
 \epsfig{file=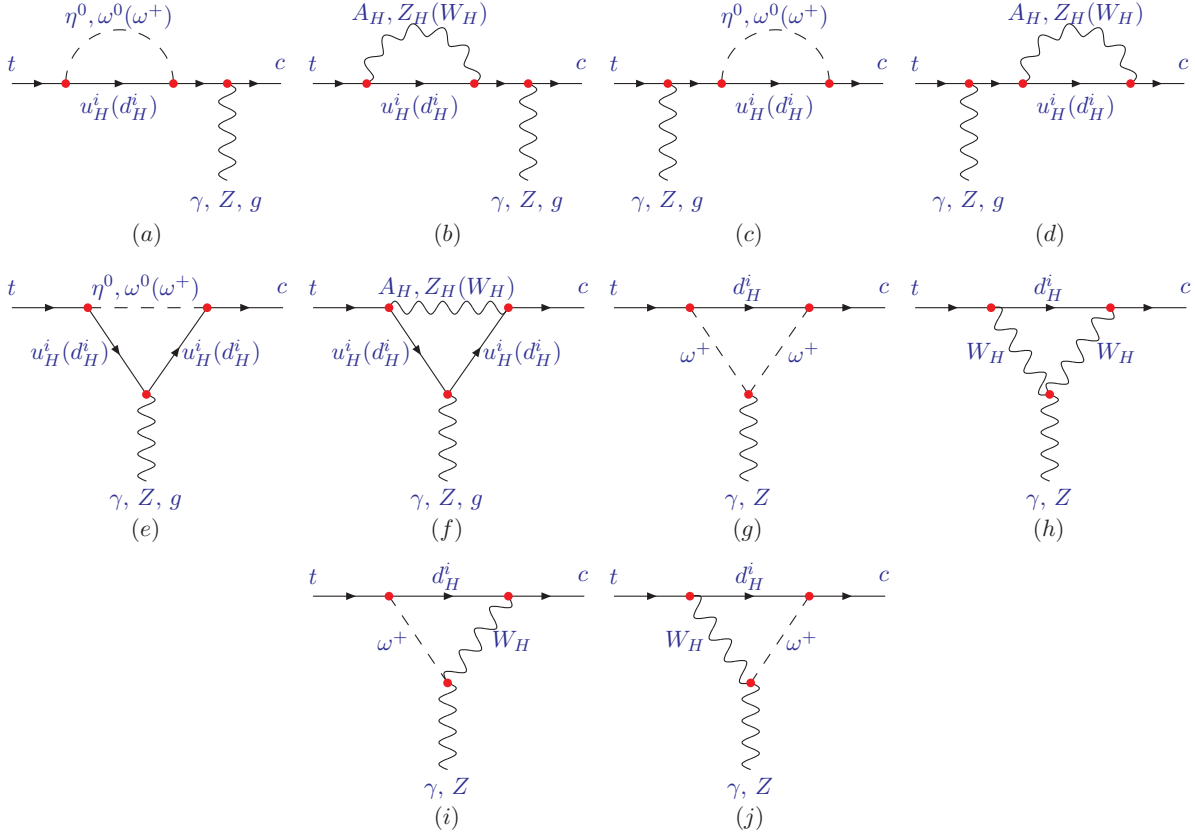,width=16cm}
\vspace*{-.5cm}
 \caption{Feynman diagrams for $t \to cV (V=g, \gamma, Z)$ at one-loop level
          in the LHT model.}
\label{fig1}
 \end{figure}

\begin{figure}[htb]
 \epsfig{file=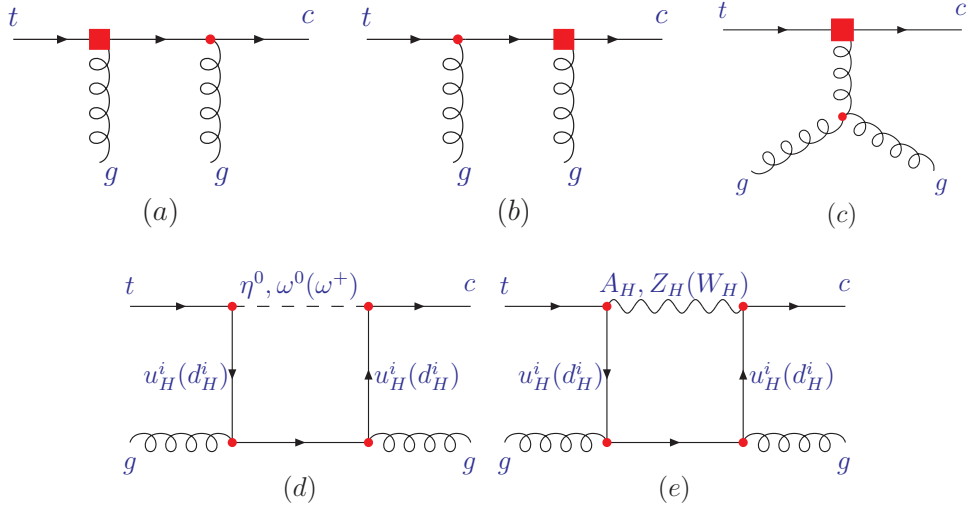,width=13cm}
\vspace*{-.5cm}
 \caption{Feynman diagrams for $t \to cgg$ at one-loop level in
 the LHT model. The loop-induced $tcg$ vertex in (a-c) is shown in Fig.\ref{fig1}.}
\label{fig2}
 \end{figure}

\begin{figure}[htb]
 \epsfig{file=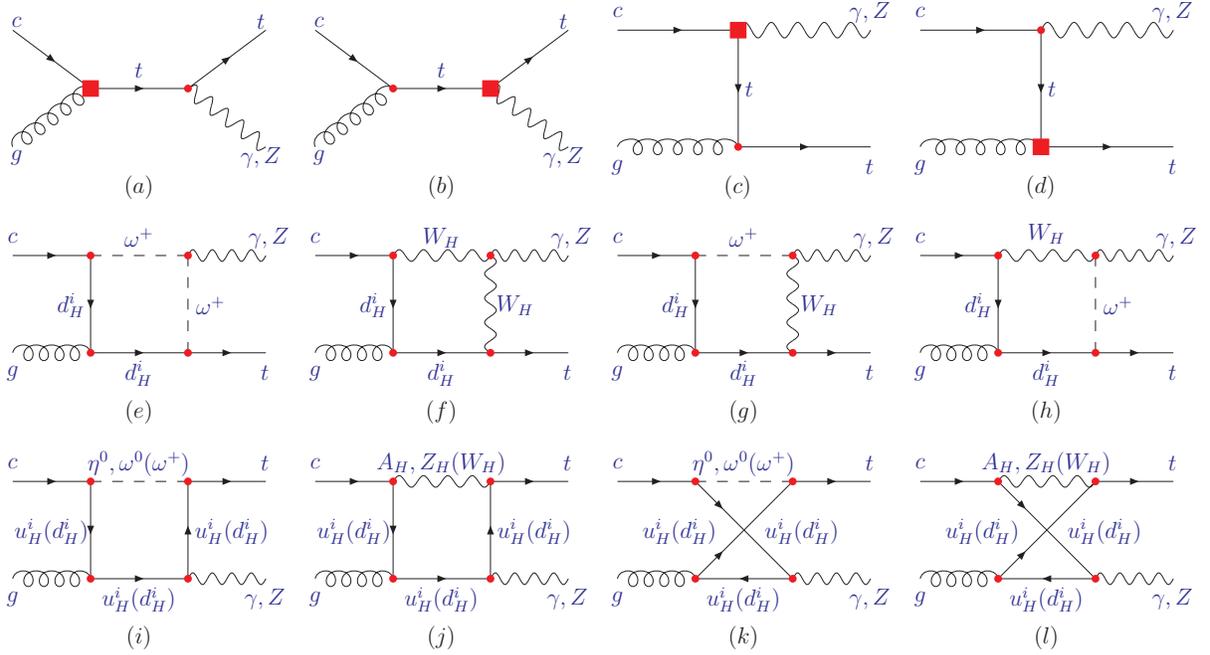,width=16cm}
\vspace*{-.5cm}
 \caption{Feynman diagrams for $cg \to tV
 (V = \gamma, Z)$  at one-loop level in the LHT model.
The loop-induced $tcV$ vertex in (a-d) is shown in Fig.\ref{fig1}.}
\label{fig3}
 \end{figure}

The LHT contributions to the FCNC top quark processes
come from the interactions between the SM
quarks and the T-odd mirror quarks, mediated by the heavy T-odd gauge
bosons or Goldstone bosons.
The relevant Feynman diagrams for the LHT contributions
are shown in Figs. \ref{fig1}-\ref{fig3}.
The Feynman diagrams for $cg \to t$, $gg \to t \bar{c}$ and $cg \to tg$
are similar to Figs. \ref{fig1}-\ref{fig3} and not plotted here.

The calculations of the loop diagrams are straightforward. Each
loop diagram is composed of some scalar loop functions
\cite{scalarloop}, which are calculated by using LOOPTOOLS
\cite{looptool}. The relevant Feynman rules can be found in
\cite{blht3}. The analytic expressions of the amplitudes for these
processes are lengthy and tedious. Here, as an example, we list
the expressions for the amplitudes of $t\to cg$ and $t\to cgg$ in
Appendix A. Note that we have checked that the divergences are
canceled at $\ord(v^2/f^2)$ for all the processes except the
channel $t \to cZ$. This so-called left-over divergence in the LHT
model was understood as the sensitivity of the decay amplitudes to
the ultraviolet completion of the theory \cite{blht3}.  In our
numerical calculations, we will follow \cite{blht3} to remove the
divergent term $1/\varepsilon$ and take the renormalization scale
$\mu=\Lambda$ with $\Lambda=4\pi f$ being the cutoff scale of the
LHT model. Note that in \cite{divcancel} the similar divergence in
the processes with down-type quarks or leptons as the external
particles can be cancelled via the modified interactions of the
up-type mirror fermions with the $Z$ boson. We checked that such a
modification cannot lead to the cancellation of the divergence in
$t \to cZ$.

In our numerical calculations we take the SM parameters as
$m_{t}=171.4$ GeV, $m_{Z}=91.187$ GeV, $m_{W}=80.425$ GeV,
$m_{c}=1.25$ GeV, $\alpha=1/128$ and $\alpha_{s}=0.107$.
The LHT parameters relevant to our study are the scale $f$,
the mirror quark masses and parameters in the matrices
$V_{H_u}$ and $V_{H_d}$.
For the scale $f$, its value may be as low as 500 GeV \cite{f500}.
For the mirror quark masses, from Eq.(\ref{eq4}) we get
$m_{u^i_{H}}=m_{d^i_{H}}$ at $\ord(v/f)$ and further we assume
\be
m_{u^1_{H}}=m_{u^2_{H}}=m_{d^1_{H}}=m_{d^2_{H}}\equiv m_{12},\ \
m_{u^3_H}=m_{d^3_H}\equiv m_3.
\ee
For  the matrices $V_{H_u}$ and $V_{H_d}$,
considering the constraints in \cite{blht1}, we follow \cite{houhs}
to consider two scenarios:
\begin{itemize}
\item[(I)]  $V_{H_d} =1$, $V_{H_u} =V_{CKM}^{\dagger}$.
In this scenario, the constraints on the mass spectrum of the mirror
fermions can be relaxed \cite{blht1}.
\begin{figure}[htb]
 \epsfig{file=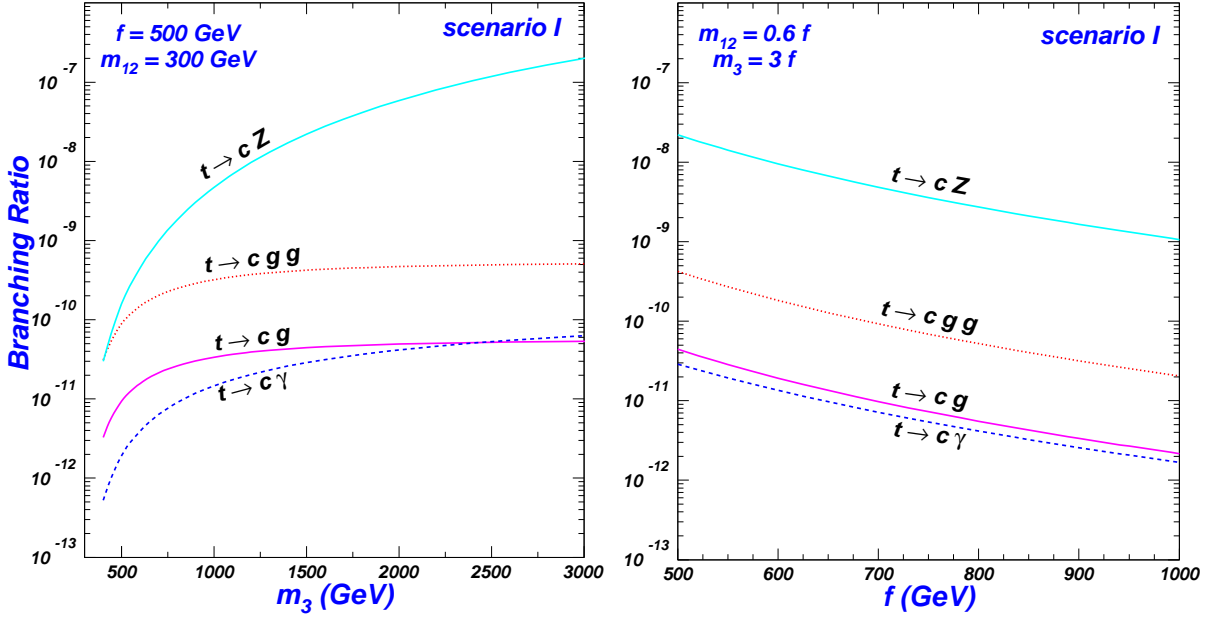,width=16cm}
\vspace*{-0.5cm}
 \caption{The branching ratios of the top-quark FCNC decays in scenario I.}
 \label{fig4}
 \end{figure}
The decay branching ratios  in this scenario are plotted in Fig. \ref{fig4},
where we fixed $m_{12}=300$ GeV and $f=500$ GeV for the
left frame while for the right frame we assumed $m_{12}=0.6 f$
and $m_3=3 f$ ( which corresponds to fixing the Yukawa couplings
$\kappa_{q^i}$ in Eq. 4).
\item[(II)]  $s_{23}^d=1/\sqrt{2}$,
             $s_{12}^d=s_{13}^d=0$,
             $\delta_{12}^d=\delta_{23}^d=\delta_{13}^d=0$.
In this scenario the $D$-meson system can give strong constraints
on the relevant parameters \cite{blht1}. Considering these constraints,
we fixed $f=1000$ GeV and $m_{12}=500$ GeV for the results shown in
the left frame of Fig. \ref{fig5}.
In the right frame of Fig. \ref{fig5} we show the results as a function
of the scale $f$ under the assumption $m_{12}=0.5 f$ and $m_3=1.2 f$.
\begin{figure}[htb]
 \epsfig{file=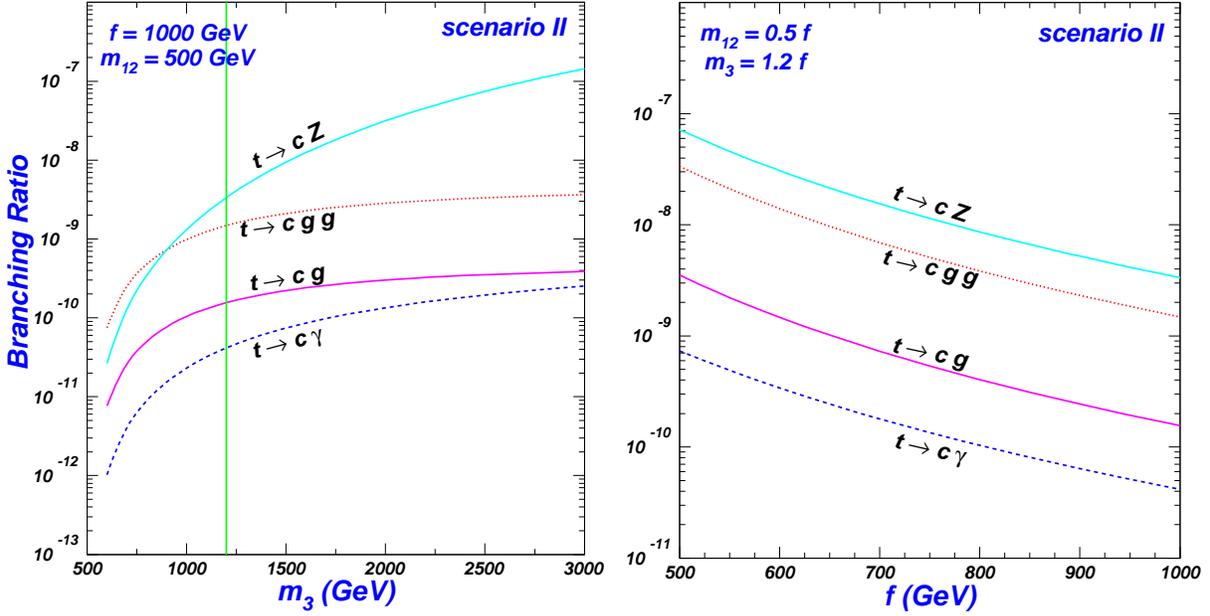,width=16cm}
\vspace*{-0.5cm}
 \caption{The branching ratios of the top-quark FCNC decays in scenario II.
 The vertical line in the left frame is the upper bound on $m_3$
 from \cite{blht1}.}
 \label{fig5}
 \end{figure}
\end{itemize}

As shown in the left frames in Figs. \ref{fig4} and \ref{fig5},
the branching ratios increase with the mass of the third
generation mirror fermions. The reason is that the decays are
enhanced by the large  mass splitting $m_3-m_{12}$, which
increases as $m_3$ gets large  since we fixed the value of
$m_{12}$. From our numerical calculation we found that the
contribution of each Feynman diagram in Fig. \ref{fig1} increases
drastically with $m_3$, but there is a strong cancellation between
different diagrams for the decays  $t\to cg, c\gamma,cgg$. For the
decay $t\to cZ$, such a cancellation is weak because of the
left-over divergence. So the enhancement with $m_3$ is rapid for
$t\to cZ$ but mild for other decay modes. As shown in the right
frames in Figs. \ref{fig4} and \ref{fig5}, the  branching ratios
drop as the scale $f$ (together with $m_{12}$ and $m_3$) gets
large, showing the decoupling behavior of the scale $f$ in the
FCNC top quark decays.

From Figs. \ref{fig4} and \ref{fig5} we see that the branching ratio of
$t \to cgg$ is larger than that of $t \to cg$. Such a feature was
also found in the SM \cite{tcvh-sm} and the minimal supersymmetric model \cite{tcv-mssm},
and the reason is explained in the literature \cite{tcv-mssm}.
Another peculiar and unexpected phenomenon is that the branching ratio of
$t \to cZ$ is the largest. This is unique to the LHT model.
The reason is that, unlike other decay modes, $t \to cZ$ is special since
it has the left-over divergence and is sensitive to the cut-off scale.

Now we turn to the top-quark FCNC productions at the LHC and present
some numerical results.
In our calculations we use CTEQ6L \cite{cteq} for parton distributions,
with the renormalization scale $\mu_R$ and factorization
scale $\mu_F$ chosen to be $\mu_R=\mu_F=m_t$.
In the following we use the parton processes to label
the corresponding hadronic processes and
all the cross sections displayed in our numerical results
are the hadronic cross sections. Also, we take into account the
charge conjugate channel for each process.

\begin{figure}[htb]
 \epsfig{file=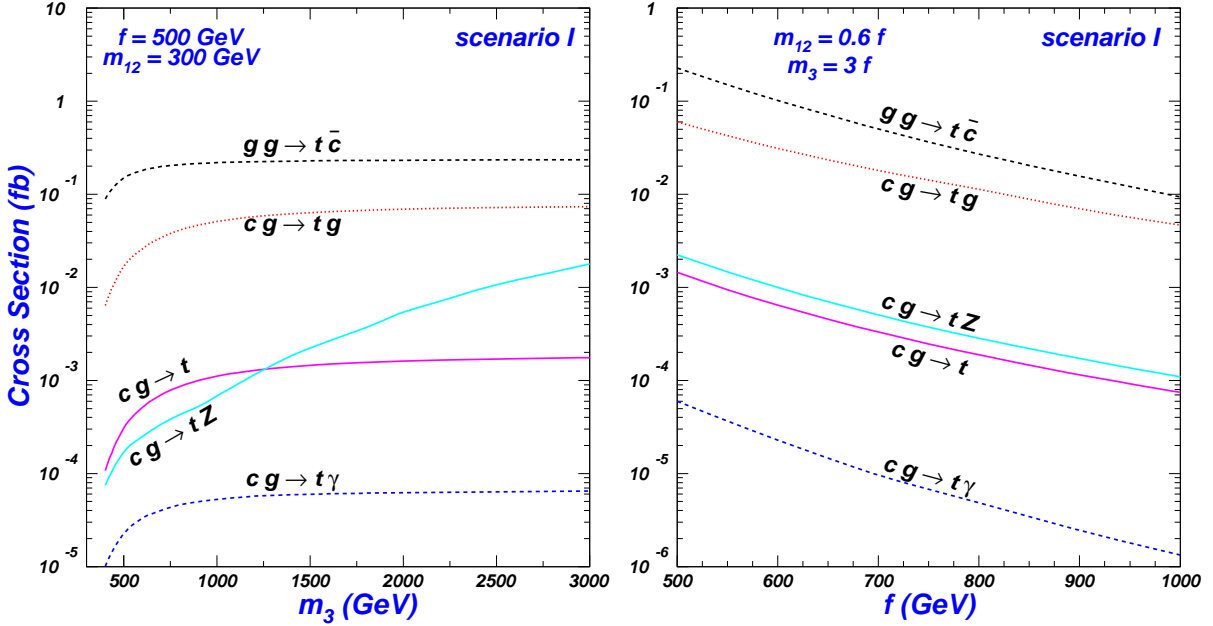,width=16cm}
\vspace*{-.8cm}
 \caption{The hadronic cross sections of FCNC top-quark
 productions in scenario I.}
 \label{fig6}
 \end{figure}

\begin{figure}[htb]
 \epsfig{file=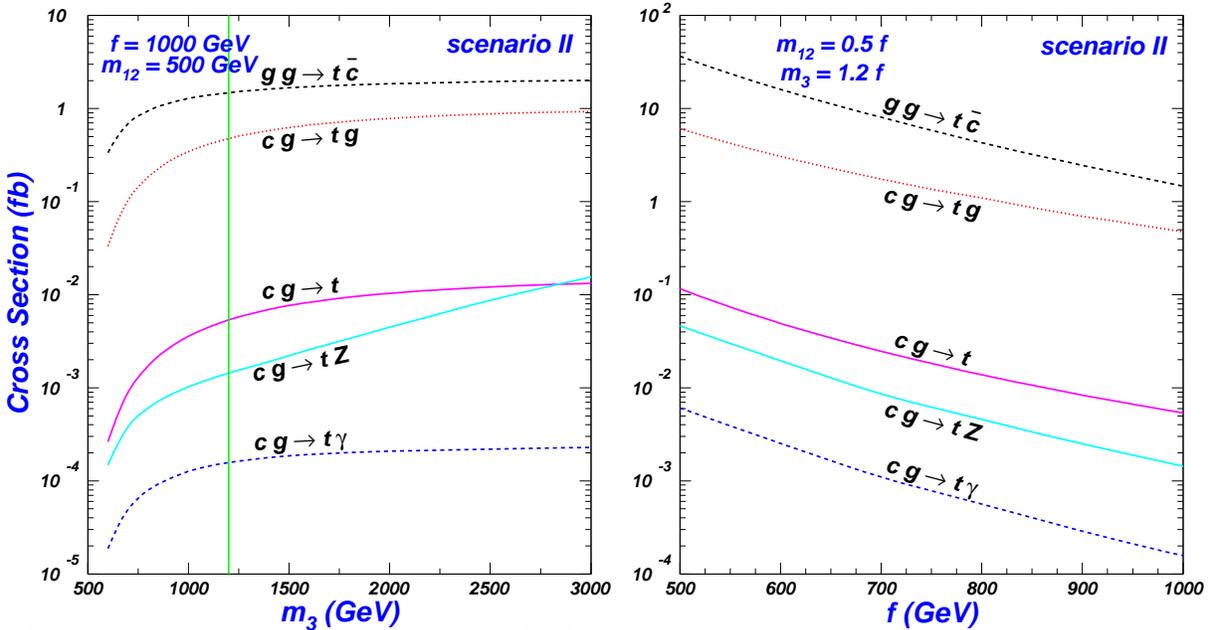,width=16cm}
\vspace*{-.8cm}
 \caption{The hadronic cross sections of FCNC top-quark
 productions in scenario II.
The vertical line in the left frame is the upper bound on $m_3$
 from \cite{blht1}.}
 \label{fig7}
 \end{figure}

The cross sections are plotted
in Figs. \ref{fig6} and \ref{fig7} for scenario I and II, respectively.
The behavior of the curves is similar to those in
Figs. \ref{fig4} and \ref{fig5}, i.e., increase with $m_3$
and decrease with the scale $f$. Also, similar to the decay
$t\to cZ$, the production rate of $cg \to t Z$ increases rapidly
with $m_3$ because of its left-over divergence at $\ord(v^2/f^2)$.

For the top FCNC decays, the LHC sensitivity with an integrated
luminosity of 100 fb$^{-1}$ is about $10^{-5}$ for  $t \to c\gamma$
and $t \to cZ$ \cite{tcrz-lhc} while for  $t \to cg$ and $t \to cgg$
the sensitivity may be much worse \cite{Aguilar-Saavedra:2004wm}.
From Figs. \ref{fig4} and \ref{fig5} we see that the decay branching
ratios are below $10^{-7}$ and thus are not accessible at the LHC.

For the top FCNC productions, the LHC sensitivity is at pb level for
$cg\to t$, $gg \to t\bar{c}$ and $cg \to tg$ while at fb level
for  $cg \to tZ$ and $cg \to t\gamma$ \cite{pptc}.
From Figs. \ref{fig6} and \ref{fig7} we see that the
the top FCNC productions in the LHT model are not accessible at the LHC.

Therefore, we conclude that although the LHT model can enhance the
top quark FCNC processes relative to the SM predictions, its
contributions are not large enough to be accessible at the LHC. This
is in contrast to the topcolor-assisted technicolor models which
give exceedingly large contributions above the LHC sensitivity
\cite{tcv-tc2}. The minimal supersymmetric model with R-parity
conservation gives quite mild contributions to these FCNC processes
of the top quark and only a couple of channels can marginally reach
the LHC sensitivity in a tiny part of the parameter space
\cite{tcv-mssm}. If R-parity is violated, then the minimal
supersymmetric model can give large contributions \cite{tcv-mssm}.
So, if the top quark properties are proved to be SM-like at the LHC
and hence the top FCNC processes are not observed, the
topcolor-assisted technicolor models and the R-parity violation in
supersymmetric models will be severely constrained, while the
R-conserving minimal supersymmetric model will be very mildly
constrained and the LHT model will not be constrained.

Note that in \cite{houhs} the LHT contributions to some FCNC top
decay processes were found to be quite large. Unfortunately, our
calculations cannot reproduce such large effects. Our results
indicate that the LHT model does not cause flavor problem for the
top quark sector.

\section{Conclusions}
In the littlest Higgs model with T-parity the
T-odd mirror quarks have flavor-changing couplings with
the Standard Model quarks and may enhance the FCNC top quark interactions.
We performed a comprehensive study for the
contributions of these mirror fermions to various top quark FCNC
decays and productions at the LHC. We found that although these
FCNC processes can be greatly enhanced by the contributions of the
mirror quarks, they are hardly accessible at the
LHC. Therefore, this model may not cause the
FCNC problem in the top quark sector if the top quark property is
proved to be SM-like at the LHC.

\section*{Acknowledgement}
This work was supported  by the National Natural
Science Foundation of China (NNSFC) under Nos. 10821504,
10725526 and 10635030.

\appendix

\section{The amplitudes of  $t \to cg$ and $t \to cgg$ in LHT}
The amplitudes for various top FCNC processes are complicated and lengthy.
Here, we only take $t \to cg$ and $t \to cgg$ for examples.
The amplitude of $t\to cg$ is given by
\be
 \mathcal{M}=\frac{i g_s T^a}{16\pi^{2}} \bar{u}_c (p_2) [(L_1 q^{\mu} + L_2 p_1^{\mu}
 + L_3 \gamma^{\mu}) P_L + (R_1 q^{\mu} + R_2 p_1^{\mu}
 + R_3 \gamma^{\mu}) P_R] u_t (p_1) \epsilon_{\mu}(q,\lambda),
\ee
where $p_1$,$p_2$, and $q$ are the real momenta of top quark, charm quark
and gluon respectively, $\epsilon_{\mu}(q,\lambda)$ is the polarization vector
of the gluon, $P_{L,R}=(1 \mp \gamma_5)/2$, $T^a$ are the generators of $SU(3)_C$.
The factors $L_i, R_i (i=1,2,3)$ are the LHT contributions from the diagrams
in Fig. \ref{fig1}, which include the scalar parts $(L_i)_S, (R_i)_S$
and vector parts $(L_i)_V, (R_i)_V$.
The scalar parts are given by
\begin{eqnarray}
(L_1)_S &=& -2 m_c b_2 a_3 (C_{21}+C_{11}) + 2 m_t a_2 b_3 (C_{23}+C_{12})
      + 2 m_f a_2 a_3 (C_{11}+C_0),\\
(L_2)_S &=& 2 m_c b_2 a_3 (C_{21}+C_{11}-C_{23}-C_{12}) + 2 m_t a_2 b_3 (C_{22}-C_{23}) \nonumber\\
&& + 2 m_f a_2 a_3 (C_{12}-C_{11}-C_0), \\
(L_3)_S &=& b_2 a_3 [m_c^2 (C_{23}+C_{12}-C_{21}-C_{11})
 + m_t^2 (C_{23}-C_{22}) + 1/2 - 2 C_{24}] \nonumber\\
&& + m_c m_t a_2 b_3 (C_{12} - C_{11})
   + m_f (m_t b_2 b_3 + m_c a_2 a_3 + m_f b_2 a_3) C_0 \nonumber\\
&& +\frac{1}{m_t^2 - m_c^2} \{b_2 a_3 [m_t^2 (B_0(p_1) + B_1(p_1))
   - m_c^2 (B_0(p_2) + B_1(p_2))] \nonumber\\
&& + a_2 b_3 m_c m_t (B_0(p_1) + B_1(p_1) - B_0(p_2) - B_1(p_2)) \nonumber\\
&& + b_2 b_3 m_f m_t (B_0(p_1) - B_0(p_2)) + a_2 a_3 m_f m_c (B_0(p_1) - B_0(p_2))\},\\
(R_1)_S &=& 2 m_t b_2 a_3 (C_{23}+C_{12}) - 2 m_c a_2 b_3 (C_{21}+C_{11}) + 2 m_f b_2 b_3 (C_{11}+C_0),\\
(R_2)_S &=& 2 m_t b_2 a_3 (C_{22}-C_{23}) + 2 m_c a_2 b_3 (C_{21}+C_{11}-C_{23}-C_{12}) \nonumber\\
&& + 2 m_f b_2 b_3 (C_{12}-C_{11}-C_0),
\end{eqnarray}
\begin{eqnarray}
(R_3)_S &=& a_2 b_3 [m_c^2 (C_{23}+C_{12}-C_{21}-C_{11})
   + m_t^2 (C_{23}-C_{22}) + 1/2 - 2 C_{24})] \nonumber\\
&& + b_2 a_3 m_c m_t (C_{12}-C_{11})
   + m_f (b_2 b_3 m_c + a_2 a_3 m_t + a_2 b_3 m_f)C_0 \nonumber\\
&& +\frac{1}{m_t^2 - m_c^2} \{a_2 b_3 [m_t^2 (B_0(p_1) + B_1(p_1))
   - m_c^2 (B_0(p_2) + B_1(p_2))] \nonumber\\
&& + b_2 a_3 m_c m_t (B_0(p_1) + B_1(p_1) - B_0(p_2) - B_1(p_2)) \nonumber\\
&& + a_2 a_3 m_f m_t (B_0(p_1) - B_0(p_2)) + b_2 b_3 m_f m_c (B_0(p_1) - B_0(p_2))\}
\end{eqnarray}
where $C_{ij}(-p_2,p_1,m_f,m_S,m_f)$, $B_{i}(p_1)(p_1,m_S,m_f)$ and $B_{i}(p_2)(p_2,m_S,m_f)$
are the loop functions \cite{scalarloop}.
The vector parts are given by
\small
\begin{eqnarray}
(L_1)_V &=& 4 m_c c_2 c_3 (C_{21}+C_{11}) ,\\
(L_2)_V &=& 4 m_c c_2 c_3 (C_{23}-C_{21}) ,\\
(L_3)_V &=& 2 c_2 c_3 [m_c^2 (C_{21} - C_{23})
  + m_t^2 (C_{22}+C_{12}-C_{23}-C_{11}) - m_f^2 C_0 + 2C_{24} - 1] \nonumber\\
&& + \frac{c_2 c_3}{m_t^2 - m_c^2} [m_c^{2}(2 B_0(p_2) + 2 B_1(p_2) - 1)
- m_t^2 (2 B_0(p_1) + 2 B_1(p_1) - 1)] ,\\
(R_1)_V &=& - 4 m_t c_2 c_3 (C_{23}+C_{11}) ,\\
(R_2)_V &=& 4 m_t c_2 c_3 (C_{23}+C_{11}-C_{22}-C_{12}) ,\\
(R_3)_V &=& 2 c_2 c_3 m_c m_t (C_{12} - C_{11})
+ \frac{2 m_c m_t c_2 c_3} {m_t^2 - m_c^2} [B_0(p_2) + B_1(p_2) - B_0(p_1) - B_1(p_1)]
\end{eqnarray}
\normalsize
with $C_{ij}(-p_2,p_1,m_f,m_V,m_f)$, $B_{i}(p_1)(p_1,m_V,m_f)$ and $B_{i}(p_2)(p_2,m_V,m_f)$.
Other relevant parameters are from
\begin{eqnarray}
S\bar{c}f&:&a_2 P_L+b_2 P_R,~~\hspace{1cm}
S\bar{f}t:a_3 P_L+b_3 P_R,\nonumber\\
V\bar{c}f&:&i\gamma^{\mu} c_2 P_L,\hspace{2cm}
V\bar{f}t:i\gamma^{\mu} c_3 P_L,
\end{eqnarray}
where $V$ represents gauge bosons and $S$ represents scalar
particles. These couplings represent the five different classes of
vertexes involved in our calculation. In each class of vertexes, the
parameters, $a_2$, $b_2$, $a_3$, $b_3$, $c_2$, $c_3$, take different
values for the every concrete coupling, respectively. The analytic
expressions of parameters are complicated at $\ord(v^2/f^2)$ and can
be found in \cite{blht3}.

Now we give the amplitude of $t \to cgg$. The expressions of Fig.\ref{fig2}(a-c)
are simple and can be obtained straightforwardly from the effective vertex
of $tcg$. For the box diagrams (d) and (e) in Fig.\ref{fig2}, their
expressions are given by
\begin{eqnarray}
 \mathcal{M}_{d} &=& -\frac{i g_s^2} {16 \pi^2} T^a_{ij} T^b_{jk}
\bar{u}_c (p_2) (a_2 P_L+b_2 P_R) S_{box} (a_3 P_L+b_3 P_R) u_t (p_1) \\
 \mathcal{M}_{e} &=& -\frac{i g_s^2} {16 \pi^2} T^a_{ij} T^b_{jk}
\bar{u}_c (p_2) \gamma^\varrho c_2 P_L S_{box} c_3 P_L \gamma_\varrho u_t (p_1)
\end{eqnarray}
where
\begin{eqnarray}
S_{box} &=& D_{\alpha\beta\gamma} \gamma^\alpha \e_slash \gamma^\beta \f_slash \gamma^\gamma
 + D_{\alpha\beta} [m_f \gamma^\alpha \e_slash \gamma^\beta \f_slash +
 \gamma^\alpha \e_slash (m_f - \q_slash) \f_slash \gamma^\beta  \nonumber \\
&&+ (m_f - \o_slash + \p_slash) \e_slash \gamma^\alpha \f_slash \gamma^\beta] +
D_{\alpha} [m_f (m_f - \o_slash + \p_slash) \e_slash \gamma^\alpha \f_slash \nonumber \\
&& + (m_f - \o_slash + \p_slash) \e_slash (m_f - \q_slash) \f_slash \gamma^\alpha +
m_f \gamma^\alpha \e_slash (m_f - \q_slash) \f_slash]  \nonumber \\
&& + D_0 m_f (m_f - \o_slash + \p_slash) \e_slash (m_f - \q_slash) \f_slash
\end{eqnarray}
with $D(-p_2,p_1,q_1,q_2,m_f,m_{S(V)},m_f,m_f)$ being 4-point loop function \cite{scalarloop},
$p_1$, $p_2$, $q_1$ and $q_2$ being respectively the momenta of top quark, charm quark and two
emitting gluons, and $\epsilon_{1}$ and $\epsilon_{2}$ are the polarization vectors of gluons.

\end{document}